\documentclass[journal]{IEEEtran}
\usepackage{fancyhdr}
\usepackage{amsmath}
\usepackage{subfigure}
\usepackage{soul}
\usepackage{extarrows}
\usepackage{color}
\usepackage{mathrsfs}
\usepackage{graphicx}
\usepackage{booktabs}
\usepackage{algorithm} 
\usepackage{algorithmic} 
\begin{document}
\bibliographystyle{IEEEtran}
\title{Joint CSIT Acquisition Based on Low-Rank Matrix Completion for FDD Massive MIMO Systems}
\author{\IEEEauthorblockN{Wenqian Shen, Linglong Dai, Byonghyo Shim, Shahid Mumtaz, and Zhaocheng Wang}
\thanks{W. Shen, L. Dai, and Z. Wang are with Tsinghua National Laboratory for Information Science and Technology (TNList), Department of Electronic Engineering, Tsinghua University, Beijing 100084 (E-mails: swq13@mails.tsinghua.edu.cn, \{daill, zcwang\}@tsinghua.edu.cn).}
\thanks{B. Shim is with Institute of New Media and Communications, School of Electrical and Computer Engineering, Seoul National University, Seoul 151-742, Korea (E-mail: bshim@snu.ac.kr).}
\thanks{S. Mumtaz is with Instituto de Telecomunica\c{c}\~{o}es (E-mail: smumtaz@av.it.pt).}
\thanks{This work was supported by the National Key Basic Research Program of China (Grant No. 2013CB329203), the National Natural Science Foundation of China (Grant Nos. 61571270 and 61201185), the Beijing Natural Science Foundation (Grant No. 4142027), and the Foundation of Shenzhen government.}\vspace{-1mm}
}
\maketitle
\begin{abstract}
Channel state information at the transmitter (CSIT) is essential for frequency-division duplexing (FDD) massive MIMO systems, but conventional solutions involve overwhelming overhead both for downlink channel training and uplink channel feedback. In this letter, we propose a joint CSIT acquisition scheme to reduce the overhead. Particularly, unlike conventional schemes where each user individually estimates its own channel and then feed it back to the base station (BS), we propose that all scheduled users directly feed back the pilot observation to the BS, and then joint CSIT recovery can be realized at the BS. We further formulate the joint CSIT recovery problem as a low-rank matrix completion problem by utilizing the low-rank property of the massive MIMO channel matrix, which is caused by the correlation among users. Finally, we propose a hybrid low-rank matrix completion algorithm based on the singular value projection to solve this problem. Simulations demonstrate that the proposed scheme can provide accurate CSIT with lower overhead than conventional schemes.

\end{abstract}

\IEEEpeerreviewmaketitle
\vspace{-2mm}
\section{Introduction}
Massive multiple-input multiple-output (MIMO) technique exploiting hundreds of antennas at the base station (BS), is one of the key enabler for future 5G wireless cellular systems.
To achieve the theoretical performance gains in massive MIMO systems, accurate channel state information at the transmitter (CSIT) is crucial \cite{LinearPrecoding}.
For CSIT acquisition, frequency-division duplexing (FDD) requires direct feedback of the CSI from the users to the BS,
but such process is unnecessary for time division duplexing (TDD) since the CSIT can be obtained from the uplink channel estimation by leveraging the channel reciprocity \cite{FullDuplex}.
While many of massive MIMO works consider the TDD mode due to this reason, FDD has many benefits over TDD (especially in delay-sensitive or traffic-symmetric applications \cite{ScalingupMIMO}) and also dominates current cellular networks.
Thus, it is of importance to come up with solutions to the CSIT acquisition problem for FDD massive MIMO systems.

Conventional CSIT acquisition for FDD MIMO systems consists of two separate steps: channel estimation in the downlink and feedback of CSI in the uplink.
First, the BS transmits orthogonal pilots in the downlink,
and each user estimates its own channel using the pilot observation.
Commonly used channel estimation algorithms include least squares (LS) and minimum mean square error (MMSE).
Then, the estimated channel is fed back to the BS via dedicated uplink channels.
Since the number of pilots grows with the number of transmit antennas at the BS, overhead of downlink pilot signaling becomes overwhelming for massive MIMO systems.
Also, the overhead of CSI feedback is a serious concern due to the same reason.
In order to address these issues,
various approaches have been proposed in recent years \cite{CloseloopTraining}-\cite{RaoJ}.
In \cite{CloseloopTraining} and \cite{DownlinkTrainingCodebook}, authors propose to reduce the downlink training ovehead by carefully designing the training pilots.
In \cite{Shim}, an approach to reduce the CSI feedback overhead when the BS antennas are highly correlated has been proposed.
In \cite{RaoJ}, an approach based on compressive sensing (CS) has been proposed to reduce both the downlink training overhead and uplink CSI feedback overhead.
While this approach is promising when the channel matrices of different users are sparse and partially share common support, such is not true when these assumptions are violated.

In this letter, we propose a joint CSIT acquisition scheme based on low-rank matrix completion for FDD massive MIMO systems.
Specifically, the BS transmits pilots for downlink channel training and the scheduled users \textit{directly} feed back the pilot observation to the BS without performing the individual channel estimation.
Then, the joint recovery of the CSI for all users is performed at the BS based on the low-rank matrix completion algorithm,
whereby the low-rank property of the massive MIMO channel matrix caused by correlation among users is exploited.
In this way, the overhead of downlink channel training as well as uplink channel feedback can be reduced, which will be verified by simulation results.

\emph{Notation}:
Lower-case and upper-case boldface letters denote vectors and matrices, respectively;
$(\cdot)^T$, $(\cdot)^H$ and $(\cdot)^{-1}$ denote the transpose, conjugate transpose, and inverse of a matrix, respectively;
$\mathbf{\Phi}^{\dagger}=\mathbf{\Phi}^{H}(\mathbf{\Phi}\mathbf{\Phi}^H)^{-1}$ is the right Moore-Penrose pseudoinverse;
$\text{rank}(\mathbf{H})$ denotes the rank of $\mathbf{H}$;
$\text{vec}(\mathbf{H})$ and $\text{unvec}(\mathbf{h})$ denote the vectorization and unvectorization, respectively;
$\otimes$ denotes the Kronecker product;
$\mathbf{I}_K$ denotes the identity matrix of size $K\times K$;
$\|\cdot\|_p$ is the $l_p$-$ \text{norm}$;
$||\mathbf{H}||_*$ is the nuclear norm denoting the sum of singular values of $\mathbf{H}$.
\section{System Model}
We consider the downlink of FDD massive MIMO system with $M$ antennas at the BS and $K$ users with single receive antenna.
The BS transmits pilots $\mathbf{\phi}_t\in\mathcal{C}^{M\times1}$ at the $t$-th channel use ($t=1,2,\cdots,T$).
At the $k$-th user, the pilot observation $\mathbf{y}_k\in\mathcal{C}^{1\times T}$ during $T$ channel uses can be expressed as
\begin{equation}
\label{eq1}
\mathbf{y}_k=\mathbf{h}_k\mathbf{\Phi}+\mathbf{n}_k,
\end{equation}
where $\mathbf{\Phi}=[\mathbf{\phi}_1, \mathbf{\phi}_2,\cdots,\mathbf{\phi}_T]$ is an $M\times T$ dimensional matrix constructed from the transmitted pilots during $T$ channel uses, $\mathbf{n}_k\in\mathcal{C}^{1\times T}$ represents the independent and identically distributed (i.i.d.) additive white Gaussian noise (AWGN) with elements having zero mean and the variance $\sigma_{n_k}^2$,
the channel vector $\mathbf{h}_k\in\mathcal{C}^{1\times M}$ between the BS and the $k$-th user is given by \cite{Tse}
\begin{equation}
\mathbf{h}_k=\sum_{p=1}^{P} g_{k,p}\mathbf{a}(\theta_{p}),
\label{eq2}
\end{equation}
where $P$ is the number of resolvable physical paths,
$g_{k,p}$ is the propagation gain of the $p$-th path,
$\theta_{p}$ is the angle-of-departure (AoD) of the $p$-th path,
and $\mathbf{a}(\theta_{p})$ is the steering vector.
In this work, we consider the typical uniform linear arrays model \cite{Tse} $\mathbf{a}(\theta_{p})=[1, e^{-j2\pi \frac{D}{\lambda}\cos(\theta_{p})},\cdots,e^{-j2\pi
\frac{D}{\lambda}(M-1)\cos(\theta_{p})}]$,
where $D$ and $\lambda$ denote the antenna spacing at the BS and carrier wavelength, respectively.
\section{Proposed Joint CSIT Acquisition Based on SVP-H Algorithm}
\subsection{Proposed Joint CSIT Acquisition Scheme}
In conventional CSIT acquisition schemes, the channel vector $\{\mathbf{h}_k\}_{k=1}^K$ of each user is estimated individually using classical algorithms such as LS or MMSE, and then the estimated CSI is fed back to the BS \cite{RaoJ}.
For example, LS algorithm generates the estimated channel vector $\hat{\mathbf{h}}_k=\mathbf{y}_k\mathbf{\Phi}^{\dagger}$.
In our work, we propose a joint CSIT acquisition scheme,
where each user directly feeds back its own pilot observation $\mathbf{y}_k$ to the BS for the joint MIMO channel recovery of all users.
The aggregate pilot observation $\mathbf{Y}=[\mathbf{y}_1^T,\mathbf{y}_2^T,\cdots,\mathbf{y}_{K}^T]^T\in\mathcal{C}^{K\times T}$ for all scheduled $K$ users can be expressed as
\begin{equation}
\mathbf{Y}=\mathbf{H}\mathbf{\Phi}+\mathbf{N},
\label{eq4}
\vspace{-0.5mm}
\end{equation}
where $\mathbf{H}=[\mathbf{h}_1^T,\mathbf{h}_2^T,\cdots,\mathbf{h}_{K}^T]^T\in\mathcal{C}^{K\times M}$ is the MIMO channel matrix to be recovered,
and $\mathbf{N}=[\mathbf{n}_1^T,\mathbf{n}_2^T,\cdots,\mathbf{n}_{K}^T]^T$ is the downlink noise matrix.

In the channel model (\ref{eq2}),
rich scattering is typically assumed at the user side, and most clusters\footnote{Cluster consists of lots of scatterers with similar delays, angle-of-arrivals, and angle-of departures \cite{ScalingupMIMO}.} around the BS are accessible for almost all users. It has been shown that a cluster seen by different users, so called ``joint clusters", introduces correlation among users even when they are geographically separated\footnote{If there exist a few ``non-joint clusters", for example, the $p^*$-th cluster around the BS, which is accessible from all users except for the $k$-th user, we can model it by setting the corresponding propagation gains $g_{k,p^*}=0$.} \cite{ScalingupMIMO}.
That is, the channel vectors associated with different users have the same steering vectors $\{\mathbf{a}(\theta_{p})\}_{p=1}^{P}$.
Thus, we have
\begin{equation}
\mathbf{H}=\mathbf{G}\mathbf{A},
\label{eq5}
\end{equation}
where $\mathbf{G}\in\mathcal{C}^{K\times P}$ with the $(k,p)$-th entry being $g_{k,p}$, and $\mathbf{A}=[\mathbf{a}(\theta_1)^T,\mathbf{a}(\theta_2)^T,\cdots,\mathbf{a}(\theta_P)^T]^T\in\mathcal{C}^{P\times M}$.
As $\text{rank}(\mathbf{H})\leq\text{min}\{\text{rank}(\mathbf{G}),\text{rank}(\mathbf{A})\}$, we have $\text{rank}(\mathbf{H})\leq\text{min}\{M,K,P\}$.
For massive MIMO systems, $M$ and $K$ are usually large but the number of resolvable paths $P$ is relatively small due to the limited number of clusters  around the BS \cite{ScalingupMIMO}, \cite{Tse},
so that $\text{rank}(\mathbf{H})\leq P$. That is, the rank of $\mathbf{H}$ of size $K\times M$ is much smaller than its dimension.
In the sequel, we call this property as ``low-rank property" of the massive MIMO channel matrix.

The pilot observation $\mathbf{Z}$ at the BS can be expressed as
\vspace{-1mm}
\begin{equation}
\mathbf{Z}=\mathbf{Q}\mathbf{Y}+\mathbf{W},
\vspace{-0.5mm}
\end{equation}
where $\mathbf{Q}\in\mathcal{C}^{M\times K}$ is the uplink Rayleigh fading channel matrix whose entries follows $\mathcal{CN}(0,1)$ \cite{AnalogFeedback},
and $\mathbf{W}\in\mathcal{C}^{M\times T}$ is the uplink noise matrix whose entries follow $\mathcal{CN}(0,\sigma_W)$.
To recover the downlink channel matrix $\mathbf{H}$ at the BS, we firstly estimate the aggregate pilot observation $\mathbf{Y}$ by \cite{AnalogFeedback}
\vspace{-1mm}
\begin{equation}
\hat{\mathbf{Y}}=(\mathbf{Q}^H\mathbf{Q})^{-1}\mathbf{Q}^H\mathbf{Z}.
\vspace{-0.5mm}
\end{equation}
Then, by exploiting the low-rank property of $\mathbf{H}$, the joint MIMO channel recovery problem at the BS can be formulated as a rank minimization problem:
\begin{equation}
\hat{\mathbf{H}}=\text{arg}\,\,\min_{\mathbf{H}}\{\text{rank}(\mathbf{H})\}, ~~\text{s.t.}~~\hat{\mathbf{Y}}=\mathbf{H}\mathbf{\Phi}.
\label{eq8}
\end{equation}
Note that this problem is non-convex and NP-hard \cite{LRMCRiemannianPursuit}.
One possible solution to avoid the computional difficulty is to use the nuclear norm minimization problem
\begin{equation}
\hat{\mathbf{H}}=\text{arg}\,\,\min_{\mathbf{H}}\{||\mathbf{H}||_*\}, ~~\text{s.t.}~~\hat{\mathbf{Y}}=\mathbf{H}\mathbf{\Phi}.
\label{eq9}
\end{equation}
Note that this problem can be solved by semidefinite programming (SDP) \cite{SDPNon-Noisy}, but the computational complexity of the solver (e.g., SeDuMi \cite{SeDuMi,Complexity}) is still high especially when the problem dimension is large in massive MIMO systems.

To alleviate the computational complexity, we need to reformulate the problem.
Firstly, we vectorize (\ref{eq4}) as
\begin{equation}
\mathbf{y}=\mathbf{\Psi}\mathbf{h}+\mathbf{n},
\label{eq7}
\end{equation}
where $\mathbf{y}=\text{vec}(\mathbf{Y})$, $\mathbf{\Psi}=\mathbf{\Phi}^T\otimes\mathbf{I}_K$, $\mathbf{h}=\text{vec}(\mathbf{H})$ and $\mathbf{n}=\text{vec}(\mathbf{N})$.
Then, the joint MIMO channel recovery problem can be reformulated as a low-rank matrix completion problem:
\begin{equation}
\hat{\mathbf{H}}=\text{arg}\,\,\min_{\mathbf{H}}\{J(\mathbf{h})=||\hat{\mathbf{y}}-\mathbf{\Psi}\mathbf{h}||_2^2\}, ~~\text{s.t.}~~\text{rank}(\mathbf{H})\leq P,
\label{eq10}
\end{equation}
where $\hat{\mathbf{y}}=\text{vec}(\hat{\mathbf{Y}})$.
Without the low-rank constraint $\text{rank}(\mathbf{H})\leq P$, it is clear that the solution to the unconstrained optimization problem $\hat{\mathbf{H}}=\text{arg}\,\min\limits_{\mathbf{H}}\{J(\mathbf{h})=||\hat{\mathbf{y}}-\mathbf{\Psi}\mathbf{h}||_2^2\}$ can be easily obtained by using the classical gradient descent algorithm or Newton's algorithm \cite{SVPG}.
However, when the low-rank constraint is added, novel algorithm must be developed to solve the constrained optimization problem (\ref{eq10}).
\subsection{SVP-H Algorithm}
The solution to (\ref{eq10}) can be obtained by using singular value projection (SVP) based algorithms \cite{SVPG} or Riemannian pursuit (RP) algorithms
\cite{LRMCRiemannianPursuit}.
In this letter, we use the modified version of the SVP-based algorithm.
For traditional SVP-based algorithms such as SVP-based gradient decent algorithm (SVP-G) and SVP-based Newton's algorithm (SVP-N), the solution satisfying the low-rank constraint can be achieved by SVP at every iteration.
In the $i$-th iteration, the current result $\mathbf{H}^{(i)}$ of linear search is projected onto a low-rank matrix $\mathbf{H}^{(i)}_q$,
which is defined as $\mathbf{H}^{(i)}_q=\text{svp}(\mathbf{H}^{(i)})=\sum_{r=1}^q\mathbf{u}_r\mathbf{\sigma}_r\mathbf{v}_r^T$,
where $\{\mathbf{\sigma}_r\}_{r=1}^q$ is the $q$ most significant singular values of $\mathbf{H}^{(i)}$. 
The resulting low-rank matrix $\mathbf{H}^{(i)}_q$ will be the starting point of a linear search for the next iteration.

\begin{algorithm}[t]
\renewcommand{\algorithmicrequire}{\textbf{Input:}}
\renewcommand\algorithmicensure {\textbf{Output:} }
\caption{The proposed SVP-H algorithm}
\begin{algorithmic}[1]
\REQUIRE ~~
$\mathbf{y}$;
$\mathbf{\Psi}$;
$q$.
\ENSURE ~~
$\mathbf{\hat{H}}$. \\
\STATE Initialization :
$\mathbf{H}^{(0)}\leftarrow \mathbf{R}$, $\mathbf{h}^{(0)}\leftarrow \text{vec}(\mathbf{H}^{(0)})$, \\ $\mathbf{H}^{(0)}_q\leftarrow \text{svp}(\mathbf{H}^{(0)})$, $\mathbf{h}_q^{(0)}\leftarrow \text{vec}(\mathbf{H}_q^{(0)})$, $i\leftarrow 1$.
\WHILE {$i\leq i_\text{max}$}
\IF {$i=1$}
\STATE $\lambda^{(i)}\leftarrow \lambda_N^{(i)}$, $\mathbf{d}^{(i)}\leftarrow \mathbf{d}_N^{(i)}$~~~~~~~~~\% SVP-N
\ELSE
\STATE $\lambda^{(i)}\leftarrow \lambda_G^{(i)}$, $\mathbf{d}^{(i)}\leftarrow \mathbf{d}_G^{(i)}$~~~~~~~~~\% SVP-G
\ENDIF
\STATE $\mathbf{h}^{(i)}\leftarrow \mathbf{h}^{(i-1)}_q + \lambda^{(i)}\mathbf{d}^{(i)}$~~~ $\mathbf{H}^{(i)}\leftarrow \text{unvec}(\mathbf{h}^{(i)})$
\STATE $\mathbf{H}^{(i)}_q\leftarrow\text{svp}(\mathbf{H}^{(i)})$ ~~~$\mathbf{h}_q^{(i)}\leftarrow \text{vec}(\mathbf{H}_q^{(i)})$
\STATE $i\leftarrow i+1$
\ENDWHILE
\RETURN $\mathbf{\hat{H}}\leftarrow \mathbf{H}^{(i)}_q$.
\end{algorithmic}
\end{algorithm}

However, as the cost function $J(\mathbf{h})$ in (\ref{eq10}) is a quadratic convex function of $\mathbf{h}$, SVP-N simply converges after one iteration (see Appendix A). Since the SVP operation is performed only once (i.e., the low-rank constraint $\text{rank}(\mathbf{H})\leq P$ will be used only once), the performance of SVP-N is generally not appealing.
On the other hand, SVP-G executes SVP in every iteration and hence a better solution can be achieved at the cost of slow convergence.
To combine the advantages of SVP-N and SVP-G, we propose the SVP-based hybrid low-rank marix completion algorithm (SVP-H) as shown in \textbf{Algorithm 1},
where SVP-N is used in the first iteration (step 4) to realize fast convergence and SVP-G is used for the rest iterations (step 6) to achieve high accuracy.
During the $i$-th iteration, the solution $\mathbf{h}^{(i)}$ is obtained through a line search along the negative gradient or Newton's direction (step 8).
After that, the unvectorized solution $\mathbf{H}^{(i)}$ of $\mathbf{h}^{(i)}$ is projected onto a low-rank matrix $\mathbf{H}^{(i)}_q$ via SVP (step 9).
The vectorized solution $\mathbf{h}^{(i)}_q$ of $\mathbf{H}^{(i)}_q$ is used as the starting point of a linear search for the next iteration.

Note that in the proposed SVP-H algorithm,
the search direction for SVP-G is the gradient $\mathbf{d}_G^{(i)}=\nabla J(\mathbf{h}_q^{(i-1)})$,
%
while the search direction for SVP-N is the Newton's direction $\mathbf{d}_N^{(i)}=\nabla^2 J(\mathbf{h}_q^{(i-1)})^{-1}\nabla J(\mathbf{h}_q^{(i-1)})$.
%
The optimal step size $\lambda^{(i)}$ is chosen to minimize $J$.
That is,
\begin{equation}
\lambda^{(i)}=\text{arg}\,\,\min_{t}\{J(\mathbf{h}_q^{(i-1)}+t\mathbf{d}^{(i)})\}.
\label{eq12}
\end{equation}
Denoting $\varphi(t)=J(\mathbf{h}_q^{(i-1)}+t\mathbf{d}^{(i)})$, then the derivative of $\varphi(t)$ is $\varphi'(t)=\nabla J(\mathbf{h}_q^{(i-1)}+t\mathbf{d}^{(i)})^T\mathbf{d}^{(i)}.$
Combining this together with $\nabla J(\mathbf{h}_q^{(i-1)}+t\mathbf{d}^{(i)})=2\mathbf{\Psi}^T\big(\mathbf{\Psi}(\mathbf{h}_q^{(i-1)}+t\mathbf{d}^{(i)})-\mathbf{y}\big)$, we have $\big(\nabla J(\mathbf{h}_q^{(i-1)})+2\mathbf{\Psi}^T\mathbf{\Psi}\mathbf{d}^{(i)}t\big)^T\mathbf{d}^{(i)}=0$,
and thus the optimal step size $\lambda^{(i)}$ is given by
\begin{equation}
\label{eq14}
\lambda^{(i)}=t=-\frac{\nabla J(\mathbf{h}_q^{(i-1)})^T\mathbf{d}^{(i)}}{{\mathbf{d}^{(i)}}^{T}(2\mathbf{\Psi}^T\mathbf{\Psi})\mathbf{d}^{(i)}}.
\end{equation}
\subsection{Complexity Analysis}
The existing algorithms to solve the SDP problem (casted from (\ref{eq9})) have high complexity $\mathcal{O}(KT)^2(K+M)^{2.5}$ \cite{SDPNon-Noisy}. If the general-purpose SDP algorithm such as SeDuMi is employed, the complexity would be burdensome \cite{Complexity}.

For the SVP-G algorithm, in each iteration, the matrix multiplication to compute the search direction $\mathbf{d}_G$ has the complexity $\mathcal{O}(KMT)$, since $\mathbf{\Psi}^T(\mathbf{\Psi}\mathbf{h}-\hat{\mathbf{y}})=
\text{vec}(\mathbf{I}_K\mathbf{H}\mathbf{\Phi}\mathbf{\Phi}^T-\mathbf{I}_K\hat{\mathbf{Y}}\mathbf{\Phi}^T)$.
The computation of the step size $\lambda_G$ is complex but we can simply assume a constant step size by allowing marginal decrease in the convergence speed \cite{SVPG}.
The SVP operation $\text{svp}(\mathbf{H})$ requires the complexity $\mathcal{O}(Mq^2)$.
Thus, the complexity of SVP-G algorithm is $\mathcal{O}((KMT+Mq^2)L)$, where $L$ is the number of iterations.
For the SVP-N algorithm, since $(\mathbf{\Psi}^T\mathbf{\Psi})^{-1}\mathbf{\Psi}^T\hat{\mathbf{y}}=
\text{vec}(\mathbf{I}_K\hat{\mathbf{Y}}\mathbf{\Phi}^T(\mathbf{\Phi}\mathbf{\Phi}^T)^{-1})$, the matrix multiplication to compute the solution $\mathbf{h}^{(i)}$ has the complexity $\mathcal{O}(KMT+M^2T+M^3)$ (see Appendix A).
There is no need to compute the optimal step size for SVP-N since the step size $\lambda_N$ for SVP-N is a constant ($\lambda_N=-1$) as shown in Appendix A.
Thus, the complexity of SVP-N is $\mathcal{O}(KMT+M^2T+M^3+Mq^2)$.
Finally, we can conclude that the proposed SVP-H algorithm has the complexity $\mathcal{O}(M^2T+M^3+KMTL+Mq^2L)$,
which is much lower than that of existing SDP algorithms.
\section{Simulation Results}
\begin{figure}
\vspace{-2mm}
\center{\includegraphics[width=0.45\textwidth]{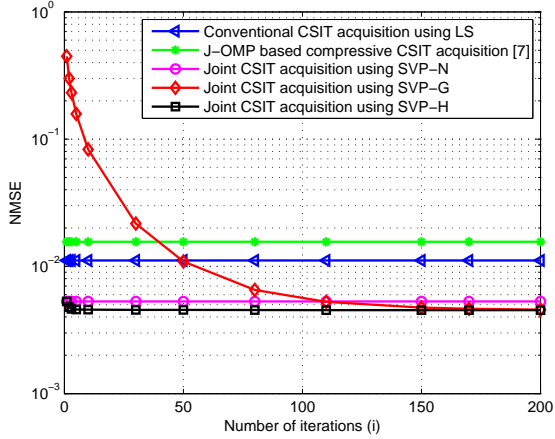}}
\vspace{-2mm}
\caption{NMSE comparison of the conventional CSIT acquisition and the proposed joint CSIT acquisition, where $T=85$ is considered.}
\label{SVP_itr}
\end{figure}

In this section, we investigate the performance of the proposed joint CSIT acquisition scheme as well as the SVP-H algorithm.
The simulation parameters are set as: $M=64$, $K=20$, $P=10$;
$\frac{D}{\lambda}=0.3$, $\theta_p=-\pi/2+\frac{p-1}{P}\pi$ \cite{ScalingupMIMO};
$i_\text{max}=250$, $q=6$.
The overhead for downlink channel training as well as uplink channel feedback are $T$ channel uses.
In Fig. \ref{SVP_itr}, we compare the normalized mean squared error (NMSE) performance of the conventional CSIT acquisition scheme and the proposed joint CSIT acquisition scheme.
Note that the NMSE of the joint orthogonal matching pursuit (J-OMP) algorithm based compressive CSIT estimation and feedback scheme proposed in \cite{RaoJ} is also presented for comparison.
The uplink channel is assumed to suffer from Rayleigh fading \cite{AnalogFeedback}, and both the downlink and uplink signal-to-noise ratio (SNR) are set to 25 dB.
As a conventional channel estimation scheme at the user side, we use the widely used LS algorithm.
In addition, the proposed joint CSIT acquisition using conventional SVP-N, SVP-G algorithms and the proposed SVP-H algorithm for joint MIMO channel recovery at the BS side are also compared in Fig. \ref{SVP_itr}.
Due to the utilization of correlations among users and the resulting low-rank property of MIMO channel matrix, it is clear that the proposed schemes using SVP-G, SVP-N, and SVP-H outperform the conventional one using LS.
The NMSE performance of J-OMP is not good because the angular-domain channel matrix in our system model does not satisfy the sparse channel assumption, and the imperfect uplink channel degrades the performance of J-OMP.
We also observe that both SVP-H and SVP-G achieve much smaller NMSE than SVP-N achieves because they repeatedly exploit the low-rank constraint as mentioned in Section III-B.
In addition, SVP-H converges faster than SVP-G due to the utilization of fast convergent SVP-N in the first iteration.
That is to say, the proposed SVP-H algorithm achieves accurate CSIT and fast convergence.
\begin{figure}
\vspace{-2mm}
\center{\includegraphics[width=0.45\textwidth]{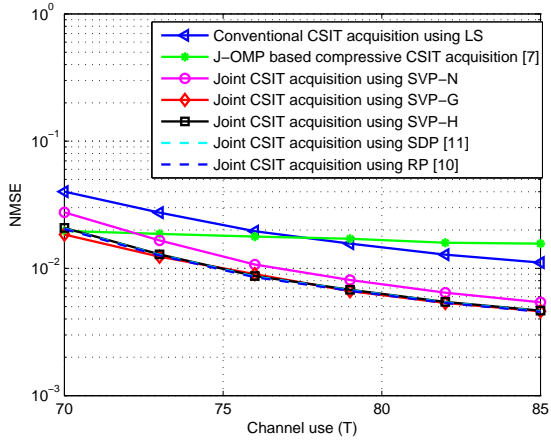}}
\vspace{-2mm}
\caption{NMSE comparison of the conventional CSIT acquisition and the proposed joint CSIT acquisition against the channel use $T$.}
\vspace{-2mm}
\label{SVP_N_C_T}
\end{figure}

We next investigate the overhead reduction of the proposed joint CSIT acquisition scheme against the channel use $T$.
Note that the NMSE performance of SDP-based method and RP algorithm have also been shown in Fig. \ref{SVP_N_C_T} for comparison.
We can observe that both approaches perform similar to the proposed SVP-H algorithm.
We can also observe that the channel use required for the proposed scheme using SVP-H is much smaller than that required for the conventional scheme.
For example, to achieve the targeted NMSE $=0.012$, the channel use required for the conventional scheme is $T=85$, while that required for the proposed scheme is $T=73$.
This clearly indicates that the proposed scheme can reduce the overhead of downlink channel training and uplink channel feedback.
\section{Conclusions}
In this letter, we investigate a novel CSIT acquisition scheme for FDD massive MIMO systems by exploiting the property that the channel matrix of massive MIMO system has low-rank structure.
Using this property, we formulate the joint CSIT acquisition scheme as a low-rank matrix completion problem.
Simulations have verified that the proposed SVP-H algorithm can achieve accurate CSIT with fast convergence.

\section{Appendix A}
\vspace{2mm}
By substituting $\nabla^2 J(\mathbf{h}_q^{(i-1)})=2\mathbf{\Psi}^T\mathbf{\Psi}$ into (\ref{eq14}),
we can compute the step size of SVP-N
\begin{align}
&\lambda^{(i)}_N=  \nonumber\\
\vspace{-2mm}&\!-\!\frac{\nabla J(\mathbf{h}_q^{(i-\!1)})^T(2\mathbf{\Psi}^T\mathbf{\Psi})^{\!-\!1}\nabla J(\mathbf{h}_q^{(i-\!1)}))}{((2\mathbf{\Psi}^T\mathbf{\Psi})^{\!-\!1}\nabla J(\mathbf{h}_q^{(i-\!1)}))^T(2\mathbf{\Psi}^T\mathbf{\Psi})(2\mathbf{\Psi}^T\mathbf{\Psi})^{\!-1\!}\nabla J(\mathbf{h}_q^{(i-\!1)})} \nonumber\\
&=-1.
\label{eq17}
\end{align}
The solution in the $i$-th iteration is given by
\begin{align}
\label{eq18}
\mathbf{h}^{(i)}&=\mathbf{h}^{(i-1)}_q-(2\mathbf{\Psi}^T\mathbf{\Psi})^{-1}2\mathbf{\Psi}^T(\mathbf{\Psi}\mathbf{h}_q^{(i-1)}-\hat{\mathbf{y}})\nonumber\\
&=(\mathbf{\Psi}^T\mathbf{\Psi})^{-1}\mathbf{\Psi}^T\hat{\mathbf{y}},
\end{align}
which is a constant vector and independent of the iteration index $i$. Thus, SVP-N algorithm obtains the solution after the first iteration.

\end{document}